\documentclass[11pt,fleqn]{article} %[twocolumn]
\usepackage{graphicx}
\usepackage{textcomp} % for \textinterrobang
\usepackage{epstopdf}
\usepackage{latexsym}
\usepackage{amsmath}
\usepackage{amssymb}
\usepackage{bm}
\usepackage[mathcal]{euscript}
\usepackage{slashed}

\usepackage[top=35mm, bottom=30mm, left=25mm, right=25mm]{geometry}
\numberwithin{equation}{section}

\usepackage{marvosym}

\usepackage{indentfirst}
\newcommand\blfootnote[1]{
  \begingroup
  \renewcommand\thefootnote{}\footnote{#1}
  \addtocounter{footnote}{-1}
  \endgroup
}

\usepackage{hyperref}
\hypersetup{
        unicode,	% use with \texorpdfstring
        colorlinks,
        citecolor=blue,% filecolor=black,%
        linkcolor=blue, % urlcolor=black,% pdftex
        urlcolor=red,
        bookmarksopen=true,
        bookmarksopenlevel=\maxdimen,
      }

\def\gl#1#2{\ifmmode \mathrm{GL}(#1; {\bf #2}) \else $\mathrm{GL}(#1; {\bf #2})$\fi}
\def\sl#1#2{\ifmmode \mathrm{SL}(#1; {\bf #2}) \else $\mathrm{SL}(#1; {\bf #2})$\fi}
\def\so#1{\ifmmode \mathrm{SO}({#1}) \else $\mathrm{SO}(#1)$\fi}

\def\sp#1#2{\ifmmode \mathrm{Sp}(#1; {\bf #2}) \else $\mathrm{Sp}(#1; {\bf #2})$\fi}
\def\usp#1{\ifmmode \mathrm{USp}(#1) \else $\mathrm{USp}(#1)$\fi}
\def\spin#1{\ifmmode \mathrm{Spin}(#1) \else $\mathrm{Spin}(#1)$\fi}
\def\su#1{\ifmmode \mathrm{SU}({#1}) \else $\mathrm{SU}(#1)$\fi}
\def\double #1{#1{\hbox{\kern-2pt $#1$}}}
\def\ket#1{\left| #1\right>}

\mathcode`\*="702A                  % * now always complex conjugate
\def\half{{\textstyle{1\over{\raise.1ex\hbox{$\scriptstyle{2}$}}}}}

\def \p{\partial}
\def \a{\alpha}
\def\ah{\hat\alpha}
\def \b{\beta}
\def\bh{\hat\beta}

\def \d{\delta}

\def \g{\gamma}
\def\gh{\hat\g}
\def \l{\lambda}

\def \o{\omega}

\def \O{\Omega}
\def \t{\theta}
\def\th{\hat\theta}

\def \Intz{{\int\! {d}^2z\,}}

\def\k{\kappa}
\def\kh{\hat\kappa}
\def\ve{\varepsilon}
\def\veh{\hat\varepsilon}
\def\s{\sigma}

\begin{document}

%%%%% preprint number goes here
\begin{flushright}
\makebox[0pt][b]{}
\end{flushright}

%%%%%%% title and authors
\vspace{40pt}
\begin{center}
{\LARGE Ambitwistor superstring in the Green-Schwarz formulation}

\vspace{40pt}
Osvaldo Chand\'ia${}^\clubsuit$ and Brenno Carlini Vallilo${}^{\spadesuit}$
\vspace{40pt}

{\em $\clubsuit$ Departamento de Ciencias, Facultad de Artes Liberales, Universidad Adolfo Ib\'a\~nez\\ Diagonal Las Torres 2640, Pe\~nalol\'en, Santiago, Chile  }\\
\vspace{40pt}
{\em $\spadesuit$ Departamento de Ciencias F\'{\i}sicas, Universidad Andres Bello\\ Sazi\'e 2212,  Santiago, Chile  }\\

\vspace{60pt}
{\bf Abstract}
\end{center}
In this paper we construct the ambitwistor superstring in the
Green-Schwarz formulation. The model is obtained from the related pure
spinor version. We show that the spectrum contains only ten
dimensional supergravity and that kappa symmetry in a curved
background implies some of the standard constraints.

%%%% email addresses
\blfootnote{\\
${}^\clubsuit$  \href{mailto:ochandiaq@gmail.com}{ochandiaq@gmail.com}\\
${}^{\spadesuit}$ \href{mailto:vallilo@gmail.com}{vallilo@gmail.com} }

\setcounter{page}0
\thispagestyle{empty}

\newpage

\tableofcontents

\parskip = 0.1in
\section{Introduction}

Perturbative superstring theory determines not only physical states
but also their interactions. These include supergravity interactions,
gauge interactions and a tower of very massive states in ten
dimensions. The presence of this infinite set of states complicates
calculations of scattering amplitudes but it also provides a control
of the quantum aspects of the theory. One could eliminate the massive
states in the limit $\alpha'\to 0$. In this way, the spectrum obtained
in perturbation quantization is the massless states only. The
world-sheet description of these class of strings has been recently
introduced in \cite{Mason:2013sva}. This new string can be seen as the
$\alpha'\to 0$ limit of the usual string. Note that they provide the scattering amplitudes proposed by Cachazo, He and Ye (CHY) \cite{Cachazo:2013gna} \cite{Cachazo:2013hca} \cite{Cachazo:2013iea} in gauge and gravity theories. A manifest space-time supersymmetric extension was constructed in \cite{Berkovits:2013xba} using the pure spinor formalism where the CHY scattering amplitudes were generalized to ten-dimensional superspace. Since supergravity theories are not exact quantum theories, it is expected that world-sheet loops contributions to scattering amplitudes should modify them. In our case, the modification should be done on type II supergravity. A one-loop analysis was done in \cite{Adamo:2013tsa} for the NS-NS part of the spectrum. It would be interesting to determine the RR contributions by using our pure spinor formulation.

A natural next step is to consider the string moving in a curved
space. In the RNS case it was done in  \cite{Adamo:2014wea} where the
superdiffemorphism algebra is satisfied if the background NS-NS fields
satisfy the supergravity equations of motion in ten
dimensions. Their result is exact for all orders in worldsheet
perturbation theory. The RR contributions can be naturally obtained using the pure spinor formalism. The coupling to a curved background was studied in \cite{Chandia:2015sfa}, where the BRST invariance of the particle-like hamiltonian of the system helps to put the background superfields on-shell. It was noted that the system has an additional symmetry whose BRST invariance implies the so-called nilpotency constraints in \cite{Berkovits:2001ue}  involving $H=dB$, where $B$ is the super two-form of type II supergravity in ten dimensions. It is natural to consider the generator of this symmetry as part of the BRST charge. This was done in \cite{Chandia:2015xfa} where
the nilpotency of the new BRST charge and the BRST invariance of the particle-like hamiltonian determine the type II supergravity constraints in ten dimensions of \cite{Berkovits:2001ue}. In this paper, we obtain the Green-Schwarz superstring for the pure spinor string of \cite{Chandia:2015xfa}.

This paper is organized as follows. In section 2 we define the GS
ambitwistor superstring in flat superspace. We show that the action is
invariant under space-time supersymmetry transformations as well as
under kappa symmetry. In this case, the necessary Virasoro-like
constraints are determined. In section 3, we relate the GS superstring
to the pure spinor ambitwistor string in flat space. The relation
allows to obtain the particle-like hamiltonian of the pure spinor
ambitwistor superstring from the Virasoro-like constraints of the GS
ambitwistor superstring. In section 4 the spectrum of the GS
ambitwistor superstring is computed using the light-cone gauge
quantization. The spectrum is described by the linearization of the type II supergravity fields, as expected. It would be interesting to apply the well known light-cone methods to
compute scattering amplitudes in this string and see if they match
with the ones obtained from the Mason-Skinner string and the pure
spinor version. We conclude in section 5 by studying the GS action in a curved superspace. The action turns out to be invariant under kappa symmetries if the constraints of type II supergravity are imposed.

\section{Flat superspace formulation}

We construct an action for the ambitwistor in Green-Schwarz formalism. We find the $\k$ symmetry of the theory and the Virasoro-like constraints. The world-sheet fields are the conjugate pairs $(X^m, P_m)$ and the super coordinates $(\t, \th)$ which describe the type II superspace in ten dimensions. As it was shown in \cite{Chandia:2015xfa} for the pure spinor case, the supersymmetry transformations for these variables are
\begin{align}
\delta \theta^\alpha =\varepsilon^\alpha,\quad
\delta \hat\theta^{\hat\alpha} = \hat\varepsilon^{\hat\a},\quad
\d X^m = -\frac{\rm i}{2} (\ve\g^m\t) - \frac{\rm i}{2} (\veh\g^m\th),\quad
\d P_m = \frac{\rm i}{2} \p (\ve\g_m\t - \veh\g_m\th) .
\label{susy}
\end{align}
Note that these transformations satisfy the usual super Poincare algebra, that is $[\d_1 , \d_2]$ leaves $(\t,\th,P)$ invariant and translates $X^m$ in an amount proportional to $\ve_1\g^m\ve_2+\veh_1\g^m\veh_2$.

It is useful to define invariant fields under supersymmetry to get invariant objects like the action. We define the combinations
\begin{align}
\Pi^m = \p X^m + \frac{\rm i}{2} \t\g^m\p\t + \frac{\rm i}{2} \th\g^m\p\th,\quad
\bar\Pi^m = \bar\p X^m + \frac{\rm i}{2} \t\g^m\bar\p\t + \frac{\rm i}{2} \th\g^m\bar\p\th ,
\end{align}
\begin{align}
P^m_L = P^m - \p X^m - {\rm i} \t\g^m\p\t,\quad
P^m_R = P^m + \p X^m + {\rm i} \th\g^m\p\th ,
\end{align}
which are invariant under supersymmetry. The combinations $P_{L,R}$ are not independent, they satisfy $P^m_R-P^m_L=2\Pi^m$. Note that $(\p\t, \bar\p\t, \p\th,\bar\p\th)$ are also invariant under supersymmetry. The Green-Schwarz action is written in terms of these supersymmetry invariant objects and it is given by
\begin{align}
S_{GS} = \int d^2z ~ \frac{1}{2} ( P^m_L + P^m_R ) \bar\Pi_m +
  \frac{\rm i}{2}  (\t\g_m\p\t) \bar\Pi^m - \frac{\rm i}{2}
  (\t\g_m\bar\p\t) \Pi^m\cr  - \frac{\rm i}{2} (\hat\t\g_m\p\hat\t)
  \bar\Pi^m + \frac{\rm i}{2} (\hat\t\g_m\bar\p\hat\t) \Pi^m
  + \frac14 (\t\g^m\p\t) (\th\g_m\bar\p\th) - \frac14 (\t\g^m\bar\p\t) (\th\g_m\p\th) .
\label{action}
\end{align}
We now verify that this action is invariant under (\ref{susy}). Consider first the term involving $\ve$ in the variation of the action. This variation becomes proportional to
\begin{align}
\int d^2z~(\ve\g^m\p\t)(\t\g^m\bar\p\t)-(\ve\g^m\bar\p\t)(\t\g^m\p\t) .
\label{susyGS}
\end{align}
On the one hand, this expression, after using the Fierz identity $\g^m_{\a(\b}(\g_m)_{\g\d)}=0$, is equal to
\begin{align}
\int d^2z~(\ve\g^m\t)(\p\t\g^m\bar\p\t) .
\end{align}
On the other hand, integrating by parts (\ref{susyGS}) is equal to
\begin{align}
-2\int d^2z~(\ve\g^m\t)(\p\t\g^m\bar\p\t) .
\end{align}
Therefore, (\ref{susyGS}) vanishes (up to a total derivative). A similar implication is obtained for the terms involving $\veh$. Then, the action (\ref{action}) is invariant under space-time supersymmetry.

The equations of motion derived from the action (\ref{susyGS}) are
\begin{align}
\bar\Pi^m=0,\quad \bar\p P_m -\frac12 \p(\t\g_m\bar\p\t)+\frac12 \p(\th\g_m\bar\p\th)=0 ,
\label{eq1}
\end{align}
\begin{align}
P^m_L(\g_m\bar\p\t)_\a=0,\quad P^m_R(\g_m\bar\p\th)_{\ah} = 0.
\label{eq2}
\end{align}
These equations are easily solved in the light-cone gauge, as we describe below, and the world-sheet fields are holomorphic.

We now verify that this action is invariant under $\k$-symmetry and find the Virasoro-like constraints for (\ref{action}). The $\k$-transformations are defined by
\begin{align}
\d \t^\a = P^m_L (\k\g_m)^\a,\quad \d \th^{\ah} = P^m_R (\kh\g_m)^{\ah} ,
\label{dkp}
\end{align}

\begin{align}
\d X^m =  \frac{\rm i}{2} \d\t \g^m \t + \frac{\rm i}{2} \d\th \g^m \th,\quad \d P_m = -\frac{\rm i}{2} \p ( \d\t \g_m \t - \d\th\g_m\th ) .
\end{align}
As usual, the $\k$ transformations for the $X$ field has the opposite sign of the supersymmetry transformation. This property is also true for the $P$ field. We can obtain the  $\k$ transformations for $P_L, P_R$ and $\Pi$. They are,
\begin{align}
\d \Pi^m = {\rm i} \d\t \g^m \p\t + {\rm i} \d\th\g^m\p\th,\quad \d \bar\Pi^m = {\rm i} \d\t \g^m \bar\p\t + {\rm i} \d\th\g^m\bar\p\th ,
\end{align}
\begin{align}
\d P^m_L = - 2{\rm i} \d\t \g^m \p\t,\quad \d P^m_R = 2{\rm i} \d\th \g^m \p\th .
\end{align}
Under $\k$-transformations the action changes as
\begin{align}
\d S_{GS} = \int d^2z ~ {\rm i} P^m_L (\d\t \g_m \bar\p\t) + {\rm i} P^m_R (\d\th\g^m\bar\p\th) .
\label{dSGS}
\end{align}
After using (\ref{dkp}), the variation of the action will vanish if Virasoro-like constraints are imposed. Consider the first term in (\ref{dSGS}). It is equal to
\begin{align}
P^m_L P^n_L (\k\g_n\g_m\bar\p\t) = P^m_L P_{Lm} (\k_\a \bar\p\t^\a).
\end{align}
Similarly, the second term in (\ref{dSGS}) is equal to
\begin{align}
P^m_R P^n_R (\kh\g_n\g_m\bar\p\th) = P^m_R P_{Rm} (\kh_{\ah} \bar\p\th^{\ah}) .
\end{align}
Plugging these results in (\ref{dSGS}) we obtain the necessity of imposing the constraints
\begin{align}
H_L=P^m_L P_{Lm} = 0,\quad H_R=P^m_R P_{Rm} = 0 ,
\label{vir0}
\end{align}
which are the Virasoro-like constraints. These constraints can be added to the action through Lagrange multipliers. The action becomes
\begin{align}
S_{GS} + \int d^2z ~ \frac14 e_L P^m_L P_{Lm} - \frac14 e_R P^m_R P_{Rm} ,
\label{SGSeH}
\end{align}
and the Largrange multipliers transform under $\k$ transformations as
\begin{align}
\d e_L = -4 \k_\a ( \bar\p\t^\a + e_L\p\t^\a ),\quad \d e_R = 4 \kh_{\ah} ( \bar\p \th^{\ah} - e_R \p \th^{\ah} ) .
\label{dke}
\end{align}

\section{Relation to the pure spinor formalism}

It is known that there is a relation between the Green-Schwarz string and the pure spinor string \cite{Oda:2001zm}. The relation begins with the observation that the momentum conjugate of the fermionic coordinates in (\ref{action}) determines a constraint which is used to define a BRST charge. This is done with the help of the pure spinor variables \cite{Berkovits:2000fe}. The definition of the conjugate variables $p_\a, \hat{p}_{\ah}$ for $\t$ and $\th$ respectively, gives the constraints
\begin{align}
d_\a = p_\a -\frac{\rm i}{2} ( P_m -\p X_m ) (\g^m\t)_\a + {1\over4} (\t\g_m\p\t) (\g^m\t)_\a,
\label{defd}
\end{align}
\begin{align}
\hat{d}_{\ah} = \hat{p}_{\ah} -\frac{\rm i}{2}  ( P_m + \p X_m ) (\g^m\th)_{\ah} - {1\over4} (\th\g_m\p\th) (\g^m\th)_{\ah} ,
\label{defdh}
\end{align}
which are the fermionic constraints used in \cite{Chandia:2015xfa} to define the pure spinor BRST charge
\begin{align}
Q = \oint \l^\a d_\a + \hat\l^{\ah} \hat{d}_{\ah} ,
\label{brst}
\end{align}
where the pure spinors $\l$ and $\hat\l$ satisfy $\l\g^m\l=\hat\l\g^m\hat\l=0$. Clearly, this condition and the algebra of $d$ and $\hat{d}$ imply that $Q^2=0$, then this charge is used as the BRST generator in the pure spinor formalism.The BRST transformation of the fermionic coordinates are
\begin{align}
Q \t^\a = \l^\a,\quad Q \th^{\ah} = \hat\l^{\ah} .
\label{qthetas}
\end{align}
The conjugate variables of the pure spinors are $\o$ and $\hat\o$ are defined up to the gauge transformations $\d\o_\a = (\l\g^m)_\a\Lambda_m$ and $\d\hat\o_{\ah} = (\hat\l\g^m)_{\ah}\hat\Lambda_{\ah}$. They transform under (\ref{brst}) as
\begin{align}
Q \o_\a = d_\a,\quad Q \hat\o_{\ah} = \hat{d}_{\ah} .
\label{qom}
\end{align}

Oda and Tonin related the pure spinor string and the Green-Schwarz such that the difference of the corresponding actions is \cite{Oda:2001zm}
\begin{align}
Q \left( \int d^2z ~ \o_\a \bar\p \t^\a + \hat\o_{\ah} \bar\p \th^{\ah}  \right) .
\label{odatonin}
\end{align}
Note that this statement does not imply that the pure spinor and the Green-Schwarz actions are cohomologically equivalent because $Q^2$ on $\o$ and $\hat\o$ are gauge transformations.

We use this property to find the pure spinor version of (\ref{SGSeH}). The action in the pure spinor formalism is (\ref{SGSeH}) plus
\begin{align}
Q \left( \int d^2z ~ \o_\a \bar\p \t^\a + \hat\o_{\ah} \bar\p \th^{\ah} - e_L \o_\a \p \t^\a - e_R \hat\o_{\ah} \p \th^{\ah} \right) .
\label{otpure}
\end{align}

Assuming the the Lagrange multipliers $e_L$ and $e_R$ are BRST invariants, we finally find that the action is
\begin{align}
S = \int d^2z ~ P_m \bar\p X^m + p_\a \bar\p \t^\a + \hat{p}_{\ah} \bar\p \th^{\ah} + \o_\a \bar\p \l^\a + \hat\o_{\ah} \bar\p \hat\l^{\ah}  + e_L T_L + e_R T_R ,
\label{Spure}
\end{align}
where
\begin{align}
T_L = \frac14 P^m_L P_{Lm}  - d_\a \p\t^\a - \o_\a \p \l^\a,\quad T_R = - \frac14 P^m_R P_{Rm} - \hat{d}_{\ah} \p \th^{\ah} - \hat\o_{\ah} \p \hat\l^{\ah} ,
\label{Tis}
\end{align}
which correspond to the pure spinor generalization of the constraints $H_L, H_R$ in the Green-Schwarz case. Note that $T_L$ and $T_R$ are stress-like tensors which were used in \cite{Jusinskas:2016qjd} to find $b$-like ghosts. That is, there exists $b_L$ and $b_R$ that satisfy $Q_L b_L=T_L$ and $Q_R b_R=T_R$.  Note also that $T_L+T_R$ is the stress-energy tensor of (\ref{Spure}) and that
\begin{align}
{\cal H} = -T_L + T_R ,
\label{theH}
\end{align}
is a symmetry of (\ref{Spure}) discovered in \cite{Chandia:2015xfa} and that helps to find the supergravity constraints when the model of (\ref{Spure}) is coupled to a curved background. Note that the last two terms in (\ref{Spure}) are BRST trivial so the can be removed from the action. The reason for this is that both $T_L$ and $T_R$ are BRST exact \cite{Jusinskas:2016qjd}. Note also that this is not the case in the bosonic string case \cite{Mason:2013sva}. Here the corresponding Lagrange multipliers are gauge fixed and lead to $bc$ ghosts in the quantization of the model.

\section{Light-cone gauge quantization and spectrum}

The action (\ref{action}) has a large set of local symmetries that
allows us to fix some of its dynamical variables. The
kappa-symmetry (\ref{dkp}) parameters can be used to fix the fermionic
coordinates to satisfy
\begin{align}
  \gamma^+\theta=0,\quad \gamma^+\hat\theta=0.
\end{align}
The only component of $\theta\gamma^m\p\theta$ that does not vanish is
$\theta\gamma^-\p\theta=\delta_{ab}\theta^a\p\theta^b$. We will
suppress $\delta_{ab}$ from now on. The same simplification happens for
the other derivative and for the other spinor variable.

In this gauge the action simplifies to
\begin{align}\label{semiLCgs}
  S_{\rm lc} =\Intz\left(P^i\bar\p X^i +P^+\bar\p X^- +P^-\bar\p X^+
  +\frac{{\rm i}}{2} P^+_L \theta^a\bar\p\theta^a+
  \frac{{\rm i}}{2} P^+_R\hat\theta^{\hat a}\bar\p\hat\theta^{\hat a} \right).
\end{align}

Off shell this is as far as we can go. The equations of motion coming
from  (\ref{semiLCgs}) imply all world-sheet variables are
holomorphic. We now use the Virasoro-like constraints to fix
\begin{align}\label{gaugeFix}
  P_L^+ =P^+-\p X^+= k^+_L,\qquad P_R^+=P^++\p X^+=k^+_R.
\end{align}
Which means that
\begin{align}
  X^+= x^+_0 +\frac12 (k^+_R-k^+_L)(\tau-\sigma),\quad P^+=\frac12 (k^+_R+k^+_L).
\end{align}
Periodicity in $\sigma$ in the closed string implies $k^+_L=k^+_R=k^+$.

We can now use canonical quantization to find the spectrum of the
model. The commutators for the spinors are
\begin{align}
  \{\theta^a(\tau,\sigma),\theta^b(\tau,\sigma')\}=
  \frac{\pi}{k^+}\delta^{ab}\delta(\sigma-\sigma'),\quad
  \{\hat\theta^{\hat a}(\tau,\sigma),\hat\theta^{\hat b}(\tau,\sigma')\}=
  \frac{\pi}{k^+}\delta^{\hat a\hat b}\delta(\sigma-\sigma').
\end{align}
These relations imply that the modes in the expansion
\begin{align}
  \theta^a=\frac{1}{\sqrt{2k^+}}\sum_{-\infty}^\infty \theta^a_n
  e^{-{\rm i}n(\tau-\sigma)},\quad
  \hat\theta^{\hat a}=\frac{1}{\sqrt{2k^+}}\sum_{-\infty}^\infty\hat\theta^{\hat a}_n
  e^{-{\rm i}n(\tau-\sigma)}
\end{align}

satisfy
\begin{align}
  \{\theta^a_n,\theta^b_m\}=\delta^{ab}\delta_{n+m},\quad
  \{\hat\theta^{\hat a}_n,\hat\theta^{\hat b}_m\}=\delta^{\hat a\hat b}\delta_{n+m}.
\end{align}

As usual, the commutation relation of the zero modes imply the vacuum is
degenerate
\begin{align}\label{vacua}
  \ket{\dot a\hat{\dot{a}}},\quad \ket{\dot a j},\quad \ket{i\hat{\dot{a}}},
  \quad \ket{ij},
\end{align}
where the spinor zero modes act as
\begin{align}
  \theta^a_0\ket{\dot a\hat{\dot{a}}}=\sigma^{a\dot{a}}_i\ket{i\hat{\dot{a}}},\quad
  \theta^a_0 \ket{i\hat{\dot{a}}} = \sigma^{a\dot a}_i\ket{\dot{a}\hat{\dot{a}}}
\end{align}
and similarly with $\hat\theta^{\hat a}$.

Now we turn to the bosonic coordinates and their conjugate
momenta. The gauge fixing (\ref{gaugeFix}) mean we can use the
Virasoro-like constraints to find $P_R^-$ and $P_L^-$ in terms of all
other variables.
The quantization of the remaining coordinates do not follow the usual
one from the bosonic string, they are in fact more closely related to
the RNS ghosts $\beta$ and $\gamma$. The mode expansions are given by
\begin{align}
  X^i = \sum_{-\infty}^\infty x_n^i e^{-{\rm i}n(\tau-\sigma)},\qquad
  P^i =  \sum_{-\infty}^\infty p_n^i e^{-{\rm i}n(\tau-\sigma)},
\end{align}
and the commutation relations are given by
\begin{align}\label{xpcommut}
  [x^i_n, p_m^j]={\rm i}\delta^{ij}\delta_{n+m}.
\end{align}
which follow from
\begin{align}
  [X^i(\tau,\sigma),P^j(\tau,\sigma')]=\pi{\rm i}\delta^{ij}\delta(\sigma-\sigma').
\end{align}
Reality of $X$ and $P$ imply that
\begin{align}
  (x^i_n)^\dagger =x^i_{-n},\qquad (p^i_n)^\dagger =p^i_{-n}.
\end{align}

It remains to find $P^-$ and $X^-$. They they are found using the
Virasoro-like constraints
\begin{align}
  P^--\p X^- &=-\frac{1}{2k^+} P_L^iP_L^i +\theta^a\p\theta^a, \cr
  P^-+\p X^- &=-\frac{1}{2k^+} P_R^iP_R^i -\hat\theta^{\hat
               a}\p\hat\theta^{\hat a}.
\end{align}

Note that the zero mode of the left hand side of the two expressions
above is the same
\begin{align}
  p^-_0=\frac{1}{2\pi}\oint\! d\sigma \left(P^--\p X^-\right)=
       \frac{1}{2\pi}\oint\! d\sigma \left(P^-+\p X^-\right).
\end{align}

Physical states $\ket{\Psi}$ have to satisfy the condition that they
are eigen states of $p^-_0$
\begin{align}\label{massShell}
  p^-_0\ket{\Psi} = k^-\ket{\Psi}.
\end{align}

This condition fixes the mass of the state $\ket{\Psi}$ using that
\begin{align}
  p^-_0&=- \frac{1}{2k^+}\Big( p^i_0p^i_0+\sum_{n=1}^\infty\big(p^i_{-n}p^i_{n} +n^2x^i_{-n}x^i_{n}
         +{\rm i}n\left(x_{-n}^ip^i_{n}-p^i_{-n}x^i_{n}\right)+
         n\theta^a_{-n}\theta^a_{n}\big)\Big),\label{zeroHl}\\
  p^-_0&=- \frac{1}{2k^+}\Big(p^i_0p^i_0+\sum_{n=1}^\infty\big(p^i_{-n}p^i_{n} +n^2x^i_{-n}x^i_{n}
         -{\rm i}n(x_{-n}^ip^i_{n}-p^i_{-n}x^i_{n})
         -n\hat\theta^{\hat a}_{-n}\hat\theta^{\hat
         a}_{n}\big)\Big) \label{zeroHr}.
\end{align}

Note that $x^i_0$, $\theta^a_0$ and $\hat\theta^{\hat a}_0$ do
not appear in the expressions above and the normal ordering constant
from $x^i_n$ and $p^i_n$ cancels the one from $\theta^a_n$ and
$\hat\theta^{\hat a}_n$.

Let us now find what are the possible eigenstates of the two operators
on the left hand side of (\ref{zeroHl}) and (\ref{zeroHr}). We will
use $\ket{\Psi}$ to represent any of the states in (\ref{vacua}). A
state with transverse momentum $\vec{k}$ is given by
\begin{align}
  \ket{\Psi}_{\vec{k}}= e^{{\rm i} k^ix^i_0}\ket{\Psi}.
\end{align}

We have to define how the whole set of oscillators act on $\ket{\Psi}_{\vec{k}}$.
In order to do that, let us define the linear combinations
\begin{align}
  \frac{{\rm i}}{\sqrt{2}} (P^i-\p X^i) =
  \frac{{\rm i}}{\sqrt{2}} \sum_{-\infty}^\infty(p_n^i +{\rm i} n x^i_n)
  e^{-{\rm i}n(\tau-\sigma)} =\sum_{-\infty}^\infty a_n^ie^{-{\rm
  i}n(\tau-\sigma)},\\
  \frac{1}{\sqrt{2}} (P^i+\p X^i) =
  \frac{1}{\sqrt{2}}\sum_{-\infty}^\infty(p_n^i -{\rm i} n x^i_n)
  e^{-{\rm i}n(\tau-\sigma)} =\sum_{-\infty}^\infty {\bar a}_n^ie^{-{\rm
  i}n(\tau-\sigma)}.
\end{align}
Using the reality condition we see that
\begin{align}
  (a_n^i)^\dagger = -a^i_{-n},\qquad ({\bar a}_n^i)^\dagger = {\bar a}^i_{-n}.
\end{align}
The commutations relation (\ref{xpcommut}) implies
\begin{align}
  [a^i_n,{\bar a}^j_m] =0,\quad [a^i_n,a^j_m]= n\delta^{ij}\delta_{n+m},\quad
  [{\bar a}^i_n,{\bar a}^j_m]=n\delta^{ij}\delta_{n+m}.
\end{align}

The Hamiltonian of the system is
\begin{align}
  H=\frac{1}{4\pi}\oint\!d\sigma ( :H_R:-:H_L:) =
  \sum_{n=1}^\infty\Big(a^i_{-n}a^i_n + {\bar a}^i_{-n}{\bar a}^i_n +
  n\theta^a_{-n}\theta^a_{n}+n\hat\theta^{\hat a}_{-n}\hat\theta^{\hat
  a}_{n}\Big).
\end{align}
The normal ordering is defined moving all positive modes to the
right. The constant resulting from this operation cancels.
The action of the oscillators on $\ket{\Psi}_{\vec{k}}$ will be
defined such that $H$ vanishes on it
\begin{align}
  a^i_n\ket{\Psi}_{\vec{k}}={\bar a}^i_n\ket{\Psi}_{\vec{k}}=
  \theta^a_n\ket{\Psi}_{\vec{k}}=
  \hat\theta^{\hat a}_n\ket{\Psi}_{\vec{k}}=0
  \quad {\rm for}\quad n\geq 1.
\end{align}
This corresponds to imposing that all positive frequency modes kills
the vacuum, as usual. We see that modes with negative $n$ create an
excitation with energy $n$.

Now it is possible to see that the only state that satisfies
(\ref{massShell}) with both
(\ref{zeroHl}) and (\ref{zeroHr}) is $\ket{\Psi}_{\vec{k}}$ with
vanishing mass. This is the expected answer
since we already know the spectrum of both RNS and pure spinor
versions of the ambitwistor string. Although we did not compute, it
is likely that the super Poincar\'e algebra only closes when $d=10$,
just like the usual GS string.

The zero mode of the operator $k\cdot P$ is not identically zero,
but acts as zero on the
physical states. This is very similar with what happens with the
$\beta\gamma$ ghosts in RNS, where the vacuum is annihilated by
$\gamma_{\frac12}$, the zero mode of $\gamma(z)$ with NS boundary
conditions. In terms of a local operator, the the vacuum corresponds
to $\delta(\gamma(z))$.  In the same way, if a covariant quantization
were possible, the construction of the operator corresponding to
$\ket{\Psi}_{\vec{k}}$ would contain $\delta(k\cdot P(z))$.
This explains the delta function present in the vertex operator
in the Mason-Skinner string.

\section{Curved background}

The action in a curved background is given by
\begin{align}
S = \int d^2z~{\cal P}_a \bar\Pi^a - \Pi^A \bar\Pi^B B_{BA} ,
\label{scurved}
\end{align}
where $\Pi^A=\p Z^M E_M{}^A, \bar\Pi^A=\bar\p Z^M E_M{}^A$ with $Z^M$ being the superspace coordinates and $E_M{}^A$ being the super-vielbein. The flat space limit of this action becomes (\ref{action}).

The $\k$-transformations are
\begin{align}
\d Z^M E_M{}^a = 0,\quad \d Z^M E_M{}^\a = ({\cal P}_a - \Pi_a)(\g^a\k)^\a,\quad \d Z^M E_M{}^{\ah} = ({\cal P}_a + \Pi_a)(\g^a\hat\k)^\a,
\label{dzC}
\end{align}
and we determine the transformation of ${\cal P}$ by demanding the invariance of the action (\ref{scurved}) under $\k$-transformations. Let us define $\s^\a=\d Z^M E_M{}^\a$ and $\s^{\ah}=\d Z^M E_M{}^{\ah}$. The variation of the action is
\begin{align}
\d S = \int d^2z ~ [ \d{\cal P}_a \bar\Pi^a +{\cal P}_a \left( -\s^\a ( \bar\Pi^A T_{A\a}{}^a + \bar\Pi^b \O_{\a b}{}^a ) - \s^{\ah} ( \bar\Pi^A T_{A\ah}{}^a + \bar\Pi^b \O_{\ah b}{}^a ) \right) \\{\nonumber}- \s^\a \Pi^A \bar\Pi^B H_{BA\a} - \s^{\ah} \Pi^A \bar\Pi^B H_{BA\ah} ],
\end{align}
where $\O$ is the Lorentz connection, $T$ is the torsion and $H=dB$. The terms involving $\Pi^A \bar\Pi^B$ with $A$ and $B$ taking spinor values vanishes implying that
\begin{align}
H_{\a\b\g}=H_{\a\b\gh}=H_{\a\bh\gh}=H_{\ah\bh\gh}=0 .
\label{Hferms}
\end{align}
Similarly, the terms involving $\s^\a\Pi^a \bar\Pi^{\bh}, \s^{\ah}\Pi^a \bar\Pi^\b$ and $\s^\a{\cal P}_a\bar\Pi^{\bh}, \s^{\ah} {\cal P}_a \bar\Pi^\b$ vanish too. This implies that
\begin{align}
T_{\a\bh}{}^a=H_{\a\bh a}=0.
\label{THaah}
\end{align}
Using these results, the variation of the action becomes
\begin{align}
\d S = \int d^2z ~ - \s^\a ( {\cal P}^a T_{\a\b a} - \Pi^a H_{\a\b a} ) \bar\Pi^\b - \s^{\ah} ( {\cal P}^a T_{\ah\bh a} - \Pi^a H_{\ah\bh a} ) \bar\Pi^{\bh}
\label{dscurv}
\end{align}
$$+\left( \d{\cal P}_a +\s^\a ( -H_{\a\b a} \Pi^\b + (T_{\a a}{}^b - \O_{\a a}{}^b){\cal P}_b - H_{\a ab}\Pi^b ) +\s^{\ah} ( -H_{\ah\bh a} \Pi^{\bh} + (T_{\ah a}{}^b - \O_{\ah a}{}^b){\cal P}_b - H_{\ah ab}\Pi^b ) \right) \bar\Pi^a.
$$
We would like to obtain a result similar to (\ref{dSGS}) of flat space. That is, the  Virasoro-like constraints should appear. In curved space they are proportional to $({\cal P}_a \pm \Pi_a)^2$. Consider the first term in (\ref{dscurv}). Because $\s^\a$ involves $({\cal P}_a-\Pi_a)$, the Virasoro-like constraint $({\cal P}_a - \Pi_a)^2$ will appear if $H_{\a\b a}=T_{\a\b a}=(\g_a)_{\a\b}$. Similarly, for the second in (\ref{dscurv}), the  Virasoro-like constraint $({\cal P}_a + \Pi_a)^2$ will appear if $H_{\ah\bh a}=-T_{\ah\bh a}=(\g_a)_{\ah\bh}$. Then, we have the constraints for the background superfields
\begin{align}
T_{\a\b a}-H_{\a\b a}=0,\quad T_{\ah\bh a}+H_{\ah\bh a}=0 .
\label{constr}
\end{align}
We also need to cancel the second term in (\ref{dscurv}) which gives the transformation of ${\cal P}$, that is
\begin{align}
\d{\cal P}_a = \s^\a \left( (\g_a)_{\a\b}\Pi^\b - (T_{\a a}{}^b-\O_{\a a}{}^b ){\cal P}_b + H_{\a ab}\Pi^b \right)- \s^{\ah} \left( (\g_a)_{\ah\bh} \Pi^{\bh} - (T_{\ah a}{}^b-\O_{\ah a}{}^b){\cal P}_b - H_{\ah ab}\Pi^b \right).
\label{dPcurv}
\end{align}
Using this, the variation of the action is, as promised, similar to flat space case and it is equal to
\begin{align}
\d S = \int d^2z~ - ({\cal P}_a - \Pi_a) (\g^a\k)^\a ({\cal P}_b - \Pi_b) \g^b_{\a\b} \bar\Pi^\b - ({\cal P}_a + \Pi_a) (\g^a\hat\k)^{\ah} ({\cal P}_b + \Pi_b) \g^b_{\ah\bh} \bar\Pi^{\bh} ,
\label{dkSC}
\end{align}
therefore if the Virasoro-like constraints are
\begin{align}
{\cal H}_L=({\cal P}_a-\Pi_a)({\cal P}^a-\Pi^a)=0,\quad {\cal H}_R=({\cal P}_a+\Pi_a)({\cal P}^a+\Pi^a)=0,
\label{virC}
\end{align}
and the action is invariant under the $\k$-transformations of (\ref{dzC}) and (\ref{dPcurv}). Note that these Virasoro-like constraints becomes the Virasoro-like constraints of (\ref{vir0}) because, as it was shown in \cite{Chandia:2015xfa}, the field ${\cal P}$ becomes $\frac12 P_L+P_R$.

As in the flat space case, the action can include the Virasoro-like costraints through the use of Lagrange multipliers so it is equal to
\begin{align}
S+\int d^2z~\frac14 e_L{\cal H}_L - \frac14 e_R{\cal H}_R.
\label{SeH}
\end{align}
To obtain the $\k$-transformations of the Virasoro-like constraints, we need to determine the corresponding variation of ${\cal P}_a \pm \Pi_a$. It turns out that these combinations transform as
\begin{align}
\d({\cal P}_a-\Pi_a) = 2\d Z^M E_M{}^\a (\g_a)_{\a\b} \Pi^\b + (T_{\a ab}-\O_{\a ab}) \d Z^M E_M{}^\a({\cal P}^b-\Pi^b)
\label{dpm}
\end{align}
$$+  (T_{\ah ab}-\O_{\ah ab}) \d Z^M E_M{}^{\ah}({\cal P}^b-\Pi^b) ,$$

\begin{align}
\d({\cal P}_a+\Pi_a) = -2\d Z^M E_M{}^{\ah} (\g_a)_{\ah\bh} \Pi^{\bh} + (T_{\a ab}-\O_{\a ab}) \d Z^M E_M{}^\a ({\cal P}^b+\Pi^b)
\label{dpplus}
\end{align}
$$+  (T_{\ah ab}-\O_{\ah ab}) \d Z^M E_M{}^{\ah} ({\cal P}^b+\Pi^b) ,$$
if we impose the further constraints
\begin{align}
H_{\a ab}=H_{\ah ab}=0 .
\label{CHab}
\end{align}
Note that the terms involving connection does not contribute to the variation of ${\cal H}_L$ and ${\cal H}_R$ because it is anti-symmetric in the vector indices. Similarly, the torsion component has to be antisymmetric in the vector indices. That is,
\begin{align}
T_{\a (ab)}=T_{\ah (ab)}=0 .
\end{align}

The variations of the Virasoro-like constraints are
\begin{align}
\d{\cal H}_L = 4 (\k_\a \Pi^\a){\cal H}_L ,\quad \d{\cal H}_R = -4 (\hat\k_{\ah} \Pi^{\ah}){\cal H}_R ,
\label{dHs}
\end{align}
and the action is invariant under $\k$-transformations if the Lagrange multipliers transform as
\begin{align}
\d e_L=4\k_\a(\bar\Pi^\a - e_L \Pi^\a),\quad \d e_R=-4\hat\k_{\ah}(\bar\Pi^{\ah}+e_R\Pi^{\ah}),
\label{des}
\end{align}
which imply the transformations of the Lagrange multipliers (\ref{dke}) in the flat space limit.

In summary, in order to have an action with $\k$-symmetry in curved background, it is necessary to impose certain constraints on the background fields. They are
\begin{align}
H_{\a\b\g}=H_{\ah\bh\gh}=H_{A\a\bh}=T_{\a\bh}{}^a=T_{\a\b a}-H_{\a\b a}=T_{\ah\bh a}+H_{\ah\bh a}=T_{\a (ab)}=T_{\ah (ab)}=0 ,
\label{sugra}
\end{align}
and it is also necessary $T_{\a\b}{}^a=\g^a _{\a\b}$ and $T_{\ah\bh}{}^a=\g^a_{\ah\bh}$. Note that (\ref{sugra}) are part of the supergravity constraints and are not enough to put the background on-shell.

\vskip 0.2in
{\bf Acknowledgements}~
The work of B{\sc cv} is partially supported by FONDECYT
grant number 1151409 and  CONICYT grant number DPI20140115.

\appendix
{\small
\bibliographystyle{abe}
\bibliography{mybib}{}
}
\end{document}